\documentstyle[aps,preprint,epsfig]{revtex}
\begin{document}

\title{ Boosting the Curie temperature with correlations in 
diluted
magnetic semiconductors}
\author{Georges ~Bouzerar\footnote{email:bouzerar@ill.fr},  Timothy ~Ziman\footnote{Also in the CNRS,email:ziman@ill.fr} }
\address{
Institut Laue Langevin,BP 156,38042 Grenoble,France\\}
\author{Josef ~Kudrnovsk\'y\footnote{email:kudrnov@fzu.cz}}
\address{
Institute of Physics, Academy of Sciences of the Czech Republic,\\
Na Slovance 2,CZ-182 21 Prague 8, Czech Republic \\}
\date{\today}


\maketitle 

\begin{abstract}
\medskip
We present a quantitative theory for  the effects of correlated doping on 
the ferromagnetism of diluted magnetic semiconductors. It predicts that 
room temperature ferromagnetism should be
possible in homogeneous, but correlated, samples of Mn$_x$Ga$_{1-x}$As.
The theory predicts lower critical temperatures for Mn$_x$Ga$_{1-x}$N.
\end{abstract} 
\vfill\eject
Semiconductors doped with magnetic impurities offer the opportunity
to integrate magnetic and semi-conducting properties\cite{Ohno}. 
For useful devices the Curie temperature T$_c$, above which
ferromagnetism disappears, should  be 
above room temperature. Attempts to find room temperature
diluted magnetic semiconductors have been many, as there are many parameters
to explore: the choice of host semiconductor, that of the doping magnetic
impurity, the degree of compensation, and methods of preparation
and treatment of the sample. From a theoretical side, while the basic
physical mechanisms are not in dispute: RKKY-like effective interactions
mediated by both the host bands and the doping band, reliable
quantitative predictions have been lacking. The simplest RKKY approach\cite{Dietl,Jungwirth}
led to predictions of  the dependence  T$_c \propto xp^{1\over 3}$ on doping $x$ and the hole density $p$
in contradiction to 
experimental results both quantitatively and qualitatively\cite{Edmonds1}. 
This formula
fails to predict observed threshold effects: below
a critical concentration there is no ferromagnetism even at low temperatures.
In general, the predictions give unrealistically high estimates of T$_c$.
\par
Recently the reasons for this have become apparent: the
treatment of spin fluctuations and the disorder were oversimplified,  and  reliable
calculations of doping dependence are now available\cite{Bouzeraretal}. Good quantitative
estimates are in agreement with the doping dependence of well characterized
samples of Mn$_x$Ga$_{1-x}$As. The same theory
predicts much lower critical temperatures for  Mn$_x$Ga$_{1-x}$N: here the 
experimental situation is more controversial.  The calculations rely on a separation
of the calculation into two distinct steps. The  first step is to make
an {\it ab initio } estimate
of effective spin-spin interaction couplings $J_{ij}$ for different neighbors
of pairs of magnetic impurities at sites $i$ and $j$ in the doped semiconductor host. These 
couplings depend on the density of magnetic impurities and the degree
of compensation. The calculation 
must be redone for each
average doping  to take into account
local fluctuations in density, which are treated in a Coherent Potential
Approximation(CPA)\cite{Josephetal}. The second  step is  to calculate the critical temperature of the resulting dilute Heisenberg model. Treating the magnetic
spins ( the S=5/2 Mn spins) as essentially classical Heisenberg spins
has
proved adequate {\it provided} that (i) the correlations are treated with
an approximation (local Random Phase Approximation) that fully includes the effect of low frequency modes
and (ii) the disorder is treated (numerically) {\it exactly}, ie by sampling
over large samples (typically  10$^5$ host sites), rather than using an effective medium theory for the 
random Heisenberg model. This gives    numerical
predictions  for T$_c(x)$, where $x$ is the doping density, which agree with
experimental values (to be seen in Figure 1) for well annealed samples. 
\par
The success of this approach allows us to examine other material parameters,
in particular the nature of the disorder, which is treated without approximation
in the second step. We note that in the first step there is an effective
medium approximation for the effect of {\it mobile} carriers which is much more robust than in the treatment of the static magnetic ions. We shall see
that it will be modified in more general situations of disorder.
The new physical parameter which we analyze is inspired by a series
 of remarkable experiments \cite{Sooetal,Blattneretal} exploring the relationship between observed
critical temperatures of samples InAs doped with Mn, and the 
local {\it correlations} in number and distances of Mn-Mn pairs. These
correlations can be measured in films produced by different techniques
of deposition, notably Molecular Beam Epitaxy and Organo-Metallic Vapor
Phase Epitaxy (OMVPE).   Correlations were measured  by extended X-ray absorption fine structure (EXAFS),
the oscillations in cross-section for  absorption of X-rays which excite
K-shell Mn electrons to high energies. As  oscillations are
due to interference of the photo-excited electron and the potentials
of neighbouring atoms, this technique looks directly at the environment
of the Mn impurities and the conclusions are relatively model independent
\cite{SayersSternLytle,Rehretal}. The authors of \cite{Sooetal,Blattneretal} observed that high
critical temperatures ($ \approx$ 320 K) were associated  
with the occurrence 
of nearest-neighbor magnetic pairs more frequent  than one would expect
from uncorrelated substitutional disorder. At the same time  samples
were observed by transmission electron microscopy to be {\it homogeneous}
down to nanometric scales: there were not simply precipitates of 
MnAs magnetic phases, as are considered to be responsible for
samples exhibiting high 
ferromagnetic temperatures in Ga$_{(1-x)}$Mn$_x$As\cite{Ando,Hartmann}.
\par
Our theory is based on the observations of such homogeneous phases  but is rather more general and provides
a mechanism of the enhancement they observed. The new physical parameter ${\cal{P}}_r$
we shall consider is the probability of enhanced nearest-neighbor correlation
in an otherwise homogeneously disordered matrix of impurities. In the limits
 ${\cal{P}}_r  = 0 $ we have simply random site substitution of magnetic impurities on
non-magnetic sites and ${\cal{P}}_r =1$ Mn impurities are introduced in strictly correlated nearest neighbor pairs. (The choice of equivalent nearest neighbor
displacement is random). For $0 < {\cal{P}}_r < 1$ there is a partial correlation:
$(x{\cal{P}}_r /2)N$  nearest-neighbor  pairs of impurities are introduced at random,
the remaining $x(1-{\cal{P}}_r )N$ impurities are introduced singly at random sites, where
$x$ is the average impurity density, $N$ the total number of substitutional sites
in the lattice. In order to calculate T$_c(x, {\cal{P}}_r )$ with correlated disorder, we 
have to modify our calculation in two ways. The first
is simply  to generate only configurations with the required correlation in nearest neighbor occupation, without changing the exchange
couplings.
This will define a theory we shall refer to as ``Correlation B'' in the following.
The second way in which correlations modify
T$_c$ is more difficult to treat exactly. It is  that 
the local correlations should be taken into account in the initial {\it ab initio }
calculations of the effective exchange couplings.
In general we would expect the correlation
to reduce the effects of disorder on couplings compared to the 
uncorrelated case. In principle an {\it ab initio } calculation could be made,
for this treating the effects of Mn-Mn  pairs in a cluster CPA approach.
As this is rather involved, we shall instead make the following ``effective concentration Ansatz'' for calculating couplings:
at a concentration $x$ and correlation parameter ${\cal{P}}_r$ we will use exchange couplings
calculated at an effective concentration corresponding to the density of 
independent scatterers, ie $(1-{\cal{P}}_r /2)x$.
At first sight this may be surprising: we are thus varying both
average density and disorder. 
We believe that this Ansatz is accurate, however,  in the concentration
range we consider, in which the primary cause of dependence on
average concentration is the  disorder due to fluctuations in local densities of impurities. In ref\cite{Bouzeraretal} we showed that for uncorrelated
substitution at
fixed concentration, ie fixed disorder,
the Curie temperature is predicted to vary little with carrier density,
provided the density is above a threshold value.
\par
In Figure 1 we show the calculation of T$_c$ as a function of x for the perfectly correlated case ${\cal{P}}_r  = 1 $
of  Mn in  Mn$_{x}$Ga$_{1-x}$As and, for contrast, that for the uncorrelated
case ${\cal{P}}_r  = 0$ (as was presented in ref. \cite{Bouzeraretal}). We see that including correlation, but  without making the Ansatz
on effective couplings, raises the critical temperature by about 25\%. Including, via the Ansatz, effects of correlation on the magnetic couplings
leads to a much more significant increase. It  predicts  
room temperature ferromagnetism for concentrations above 6 \%.  
Note that for uncorrelated, inhomogeneous
disorder T$_c = 0$  for low concentrations $x \leq x_c\approx  1.5 \%$ as suggested by the   experiments\cite{Matsukura}. The correlations may raise
this threshold value slightly.
\par The curves in Figure 1 are for the extreme cases of independent site doping and completely correlated pair doping. For comparison we show  points taken from experiment\cite{Edmonds,EdmondsNote,Edmonds1,Matsukura,Chiba}: the comparison
to the uncorrelated calculation is appropriate as the samples produced by MBE are expected
to by random independent site substitution. 
In practice if one can enhance the correlation, as apparently was the case of InAs by using OMVPE, one would not expect perfect
correlation, so it is important to include values of ${\cal{P}}_r$. 
This can be treated 
via the Ansatz in the same way and we show the result for a concentration
of 8 \% in  Figure 2. 
One can also
study higher order clusters: for example formation of trimers. If we ignore
the renormalization of the exchange energies, this in fact {\it reduces}
the critical temperature.  By the same logic of our ``effective concentration Ansatz'' we should reduce the effective concentration by a factor 3 and this
actually increases the T$_c$. 
Clearly  more systematic studies, both experimental and theoretical, 
of higher order clustering would be desirable. 


\par
In Figure 3 we show our calculated T$_c$ as a function of 
concentration for both Mn$_{x}$Ga$_{1-x}$As and Mn$_{x}$Ga$_{1-x}$N for the correlated
case ${\cal P}_r = 1$, and using the Ansatz (Correlation A). As in the case of uncorrelated, homogeneous disorder\cite{Bouzeraretal}
 the critical temperature is consistently lower in doped GaN than GaAs, despite the larger estimated coupling
between  nearest-neighbor Mn ions.  This is  because the propagation of ordering
in the bulk
is determined by the further-than-nearest neighbor coupling which is calculated
to be much weaker in GaN than in GaAs. Thus provided that correlations
are similar, our calculations
predict that doped GaAs should be a better candidate for room-temperature effects.
\par
The practical conclusions of this theory are clear: to make a high temperature
ferromagnet, the tactic should be
\par
(i) work at  concentrations in the range x=5-10\% as concentration effects
saturate  or even decrease T$_c$ at higher values of x, because of increasing
disorder and frustration.
\par
(ii) Prepare the samples such that there are correlations in the position of 
dopants, while avoiding precipitates that weaken the desired coupling
of transport and magnetism. 
The second condition (ii) is the difficult part technologically and may of course
depend on the materials.
As we noted, correlations have been found  in homogeneous samples  by Soo et al\cite{Sooetal} in (In,Mn)As.
This is presumably because at the higher temperature of preparation
of OMVPE, the natural binding of Mn-Mn dimers enhances correlations.
In terms of control,  EXAFS\cite{Sooetal}
can  determine the value of ${\cal{P}}_r$. It would be interesting
to characterize by magneto-transport measurements the effective number of 
carriers and the resistance of the samples.

\par

Our theory shows that this enhancement of ferromagnetism by correlation
of  doping is a generic property of diluted magnetic semiconductors: no special process of interaction
is needed.  The correlation is purely local and the 
phase stays homogeneous.
Formation of interstitials may  favor high T$_c$, as suggested in refs.\cite{Sooetal}
in (In,Mn)As,  but is not necessary
in our theory. Of course the near-neighbor correlations enhanced must have
ferromagnetic couplings.
While the correlation parameter is crucial, correlation does not need
to be perfect.
It would help the fundamental understanding of 
ferromagnetism if the further-nearest-neighbor couplings could be 
measured experimentally and cluster calculations could
make more accurate theoretical predictions.
The intuitive reason for enhancement is that  correlation
favors a high level
of ferromagnetic  coupling at short distances even at low concentrations,
which more than compensates the fact that   
the correlated pairs are typically further apart. 
The detailed calculations were of course
essential to demonstrate that the local order between pairs
can propagate throughout the sample via the further neighbor
couplings.

JK acknowledges the financial support from the Grant agency of 
the Academy of Sciences of the Czech Republic (A1010203 ) and 
the Czech Science Foundation 202/04/0583)

\begin{figure}[hh]
\centerline{
\psfig{file=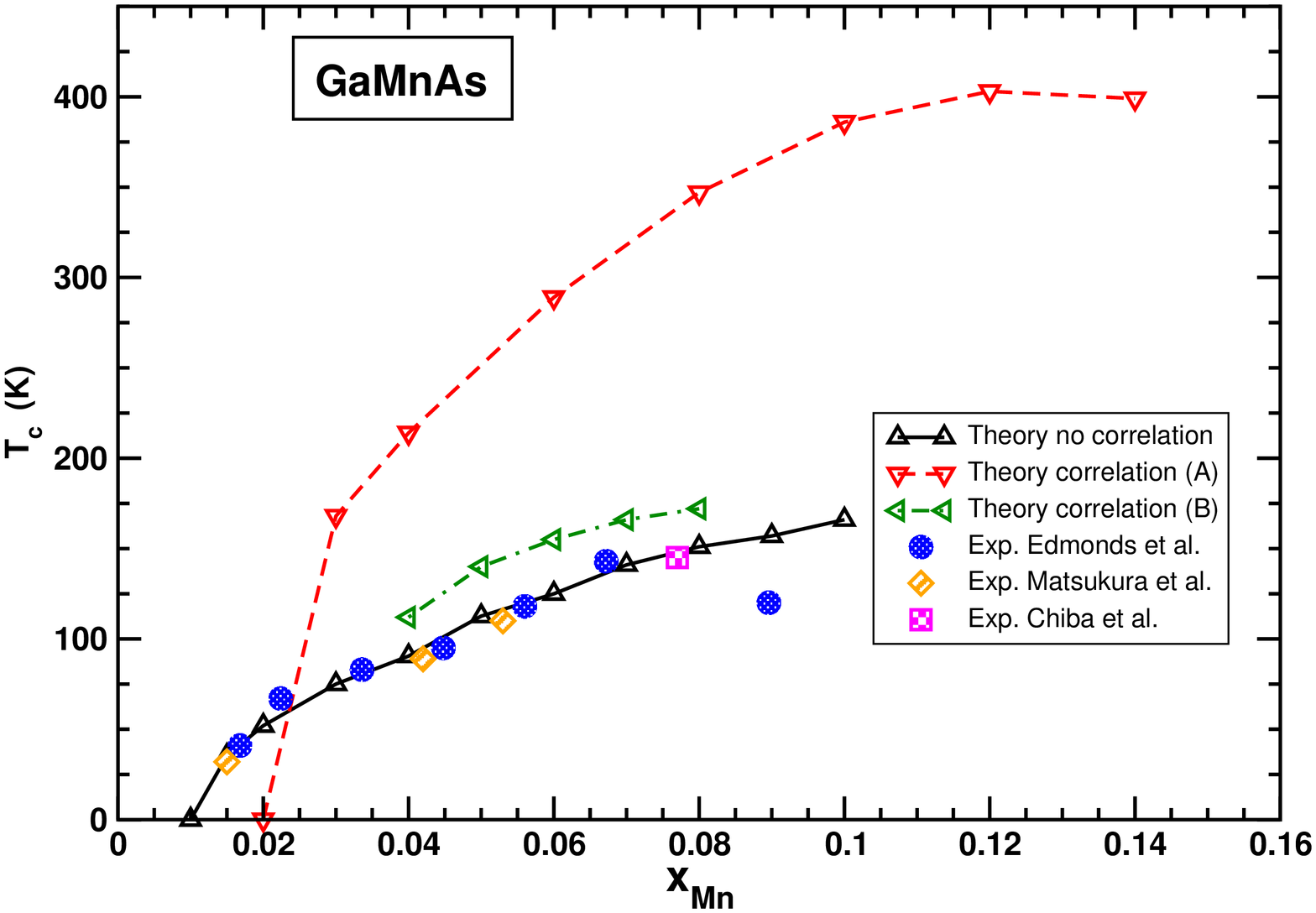,width=10cm,angle=-90}}
\caption{Calculated critical temperatures for correlated (${\cal P}_r$ =1)
and uncorrelated( ${\cal P}_r$ = 0 impurities. The correlated curve
A (resp. B) is for inclusion (resp. exclusion) of the effective concentration Ansatz, as explained in the text. 
Experimental points ( references no. 4, 13-16 )  are shown for comparison to the
uncorrelated case.}

\label{fig1}
\end{figure}

\begin{figure}[hh]
\centerline{
\psfig{file=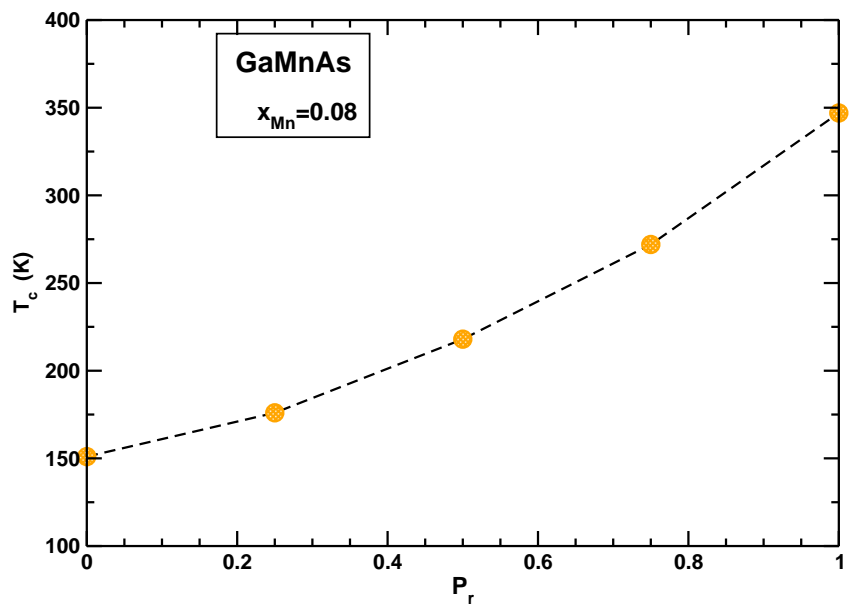,width=10cm,angle=-90}}
\caption{Enhancement of Curie temperature with correlation parameter ${\cal{P}}_r$ for fixed concentration.}\label{fig2}
\end{figure}
\vfill
\newpage
\begin{figure}[hh]
\centerline{
\psfig{file=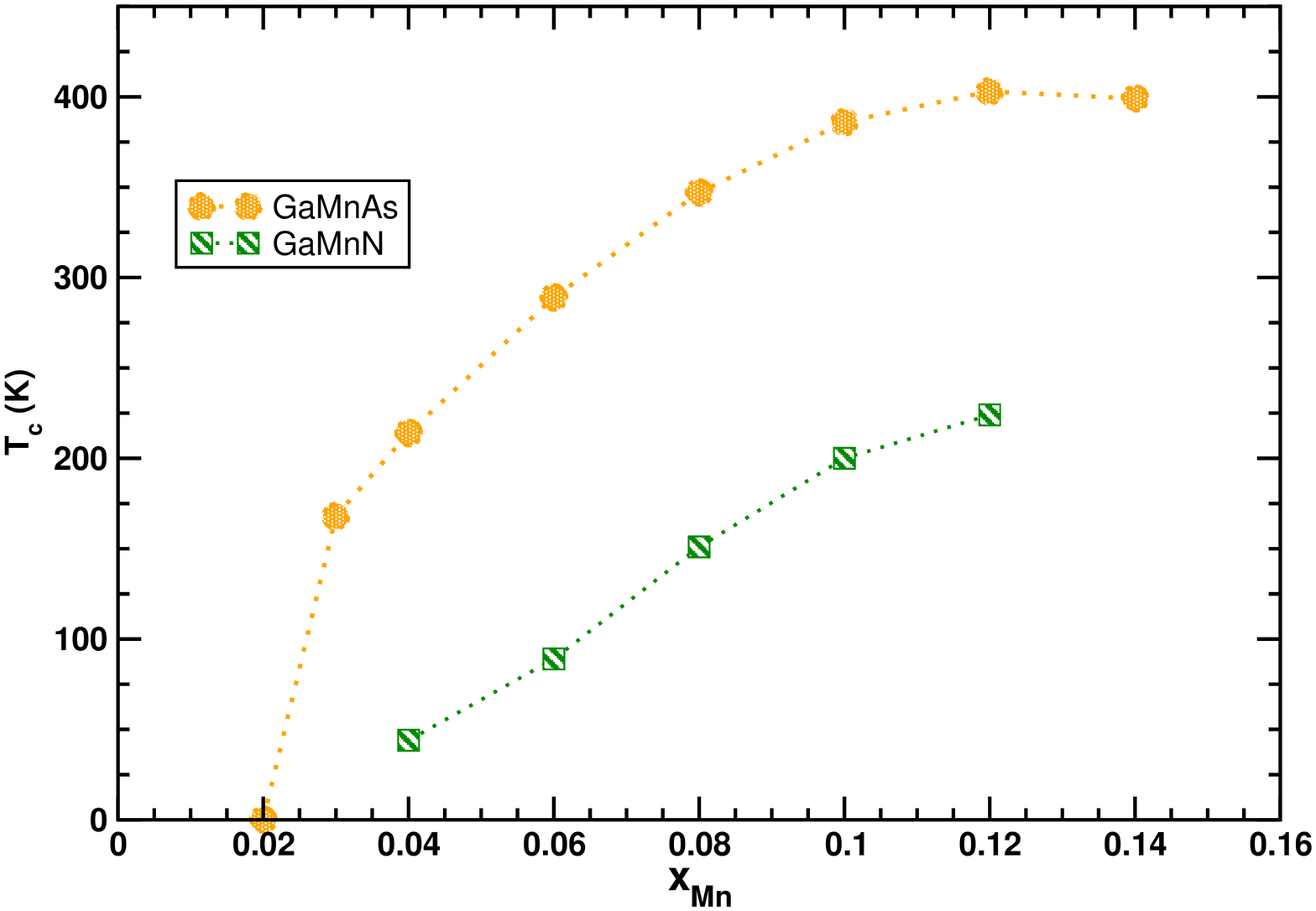,width=10cm,angle=-90}}
\caption{Predicted  T$_c$ for doped (Ga,Mn)As and (Ga,Mn)N calculated
for maximum correlation ${\cal P}_r$ = 1.}\label{fig3}
\end{figure}

\begin{references}
\bibitem{Ohno} H. Ohno, Science {\bf 281}, 951 (1998)
\bibitem{Dietl}  T. Dietl, H. Ohno, and F. Matsukura, Phys. Rev. B {\bf 63}, 195205 (2001)
\bibitem{Jungwirth}  T. Jungwirth, W. A.  Atkinson, B. H. Lee, and A. H. MacDonald, Phys. Rev. B {\bf 59}, 9818 (1999)
\bibitem{Edmonds1}  K. W. Edmonds, K. Y. Wang, R. P. Campion, A. C. Neumann, C. T. Foxon, B. L. Gallagher, and P. C. Main, 
Appl. Physics Lett. {\bf 81}, 3010 (2002)
\bibitem{Bouzeraretal} G. Bouzerar, T. Ziman, J. Kudrnovsk\'y, submitted to 
Physical Review, condmat-0405322.  
\bibitem{Josephetal}J. Kudrnovsk\'y, I. Turek, V. Drchal, F. Maca, P. Weinberger, P. Bruno  Phys. Rev. B  {\bf 69}, 115208 (2004).
\bibitem{Sooetal} Y. L. Soo, S. Kim, Y.H.Kao, A.J.Blattner, B.Wessels, S.Khalid, C.Sanchez Hanke, C.-C Kao  Appl. Phys. Lett. {\bf 84}, 481 (2004)
\bibitem{Blattneretal} A.J.Blattner,P.L. Prabhumirashi, V.P. Dravid,B.W. Wessels J. Crystal Growth {\bf 259}, 8 (2003)
\bibitem{SayersSternLytle}  D.E. Sayers, E.A. Stern and F.W. Lytle, Phys. Rev. Lett. {\bf 27}, 1204(1971).
\bibitem{Rehretal} J.J. Rehr, J. Mustre de Leon, S.I. Zabinsky, R.C. Albers, J. Am. Chem. Soc. {\bf 113}, 5135 (1991)
\bibitem{Ando} K. Ando, A. Chiba, and H. Tanoue,  Appl. Phys. Lett. {\bf 73}, 387 (1998)
\bibitem{Hartmann}  Th. Hartmann, M. Lampalzer, P.J.  Klar, W. Stolz, W. Heimbrodt,  von Nidda, H.-A. Krug, A. Loidl, L. Svistov, Physica E {\bf 13} 572 (2002) 
\bibitem{Matsukura}   F. Matsukura, H.Ohno, A. Shen, and Y. Sugawara,  Phys. Rev. B  {\bf 57}, R2037 (1998).
\bibitem{Edmonds}K. W. Edmonds et al, Phys. Rev. Lett. {\bf 92}, 037201 (2004),
\bibitem{EdmondsNote}K. W. Edmonds et al, unpublished results. We thank K. W. Edmonds for communication of these values.
\bibitem{Chiba}  D. Chiba, K. Takamura, F. Matsukura, H.Ohno,  Appl. Phys. Lett., {\bf 82}, 3020 (2003).
\end{references}
\end{document}